\documentclass{sig-alternate-05-2015}
\usepackage{mycommands}

\renewcommand{\footnotesize}{\scriptsize}  
\makeatletter
\def\@copyrightspace{\relax}
\makeatother

\begin{document}






%


\setlength{\belowdisplayskip}{2pt} \setlength{\belowdisplayshortskip}{0pt}
\setlength{\abovedisplayskip}{2pt} \setlength{\abovedisplayshortskip}{0pt}

\title{Large-Scale Query-by-Image Video Retrieval \\ Using Bloom Filters}

%
\numberofauthors{5} 
\author{
\alignauthor
Andr\'{e} Araujo\\
       \affaddr{Stanford University}\\
       \email{afaraujo@stanford.edu}
\alignauthor
Jason Chaves\\
       \affaddr{Stanford University}\\
       \email{jchaves@stanford.edu}
\alignauthor Haricharan Lakshman\\
       \affaddr{Stanford University}\\
       \email{hari.lakshman@stanford.edu}
\and  
\alignauthor Roland Angst\\
       \affaddr{Stanford University}\\
       \email{rangst@stanford.edu}
\alignauthor Bernd Girod\\
       \affaddr{Stanford University}\\
       \email{bgirod@stanford.edu}
}

\maketitle

\begin{abstract}
We consider the problem of using image queries to retrieve videos from a database.
Our focus is on large-scale applications, where it is infeasible to index each database video frame independently.
Our main contribution is a framework based on Bloom filters, which can be used to index long video segments, enabling efficient image-to-video comparisons.
Using this framework, we investigate several retrieval architectures, by considering different types of aggregation and different functions to encode visual information -- these play a crucial role in achieving high performance.
Extensive experiments show that the proposed technique improves mean average precision by $24\%$ on a public dataset, while being $4\times$ faster, compared to the previous state-of-the-art.
\end{abstract}

%
%
%

%
%

%
%
\printccsdesc
\keywords{Bloom filters, large-scale, query-by-image, video retrieval}

\section{Introduction} \label{sec:introduction}
This paper addresses the problem of searching large video databases using image queries -- a technology which enables applications such as brand monitoring and content linking.
The general task of linking images and videos has recently attracted much attention, for example for training event classifiers using web images~\cite{chen2014event,duan2012exploiting,merler2012semantic} or localizing actions/tags in videos~\cite{ballan2015data, jain2015objects2action, sun2015temporal}.
Our focus is a generic instance search problem, where no training data is available to learn classifiers for the query items.
Traditionally, this query-by-image video retrieval task has been approached by constructing systems that index each frame~\cite{le2011trecvid,meng2016object,Sivic2003} or each shot~\cite{Ballas2013,Peng2015,Zhu2014,Zhu2013,Zhu2012} in the database.
In particular, shot-level indexing achieved remarkable success in recent editions of the TRECVID Instance Search challenge~\cite{2015trecvidover}.

\begin{figure}
 \centering   
\includegraphics{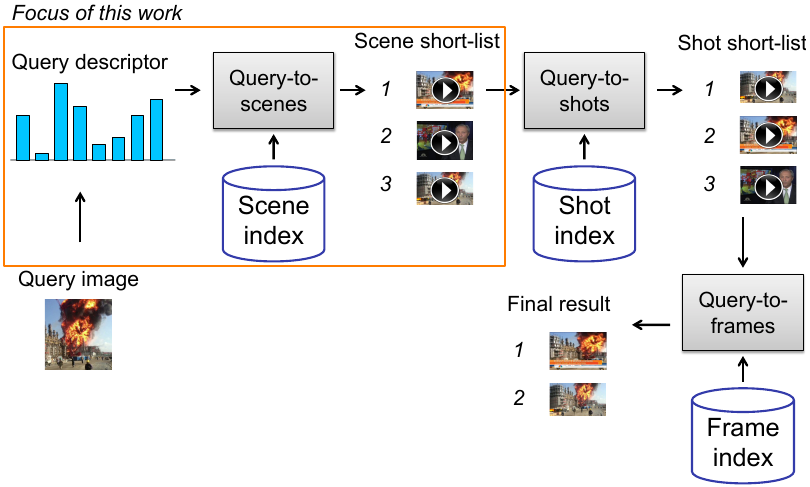}
 \caption{Block diagram of a large-scale query-by-image video retrieval system. First, a query image is represented as a descriptor which can be queried against the index of database scenes, to retrieve a short-list of scenes. Then, two refinement stages are conducted, to narrow down the matches to the shot and frame levels.
In this work, our focus is on the stage where the query image is compared directly against scenes -- this is the key to enable an efficient retrieval process.
}
  \label{fig:LSQBIVR}
\end{figure}

In contrast to most papers in this area~\cite{Ballas2013,le2011trecvid,meng2016object,Peng2015,Sivic2003,Zhu2014,Zhu2013,Zhu2012}, we consider the large-scale setting of this problem, where indexing each frame or each shot entails prohibitive latency and memory costs.
Our proposed retrieval architecture is presented in \figref{fig:LSQBIVR}: we introduce a new stage (highlighted in orange), where the query is directly compared against long video segments (scenes).
In subsequent stages of the pipeline, only a small number of scenes need to be considered for re-ranking, using shot- and frame-level information.
This architecture enables much more scalable retrieval than traditional approaches -- we demonstrate this experimentally, using three large-scale datasets.
Similar to previous work~\cite{Araujo2015ICIP,Rasheed2003,Truong2003,Yeung1998}, we define a scene as a segment of video which contains interrelated shots and represents a semantic unit for a given type of content.
We consider datasets where scenes correspond to video segments which are on average 2 to 8 \textit{minutes} long (while shots are on average 3 to 8 seconds long).

While frame- and shot-level indexing received considerable attention in previous work, only recently have researchers started to investigate methods to compare images directly against scenes.
In~\cite{Araujo2015ICIP}, Araujo \etal introduced scene-based descriptors which use Fisher vectors, enabling very compact databases and showing promising results.
However, their work presents two important disadvantages: (i) their technique obtains limited retrieval accuracy, and (ii) their large-scale system employs linear search during query time, incurring substantial retrieval latency.
Our work addresses these two aspects, to design a query-by-image video retrieval system which is scalable and achieves high retrieval accuracy.

Our main contributions are: (1) a new scene-based descriptor, which combines Fisher embedding and Bloom filters to enable up to $24\%$ more accurate and $4\times$ faster retrieval, compared to the previous state-of-the-art~\cite{Araujo2015ICIP}; (2) a study of different feature representations and hashing schemes for Bloom filters, showing that point-indexed descriptors coupled with quantizer-based hashes provide the best performance among several tested configurations; (3) large-scale experiments using three datasets (among them two newly-introduced datasets), comparing the proposed technique to the previous state-of-the-art, and to a frame-based baseline.
Contrary to~\cite{Araujo2015ICIP}, these experiments use inverted index retrieval structures for all techniques under consideration, to obtain practical large-scale systems, with reduced latency.

\section{Bloom Filters for Video \\Retrieval by Image}  \label{sec:bloom}

We are interested in efficiently retrieving database scenes that contain visual information similar to a given query image.
This problem is characterized by substantial asymmetry: while the query is a still image, database scenes are long segments of video containing diverse visual contents.
In previous work, several hashing techniques addressed the symmetric problem of image retrieval using query images \cite{Perronnin2010,wang2016learning}.
To deal with the asymmetry of our problem, we model scenes as sets and images as items.
We propose a generalization of hashing techniques, based on Bloom filters, to support efficient item-to-set comparisons.
In this section, we briefly review the concept of Bloom filters, then introduce techniques that enable efficient large-scale retrieval. 

\subsection{Review of Bloom Filters}  \label{subsec:bloom-overview}

A Bloom filter (BF) \cite{Bloom1970} is a data structure designed for set membership queries, widely used in distributed databases and networking applications -- for a review, see \cite{Broder2004}.
For a query item $q \in \myset{U}$ and a set of database items $\myset{S} \subset \myset{U}$, a BF is designed to respond to ``is $q \in \myset{S}$?''. If $q \in \myset{S}$, the answer is guaranteed to be correct (i.e., no false negatives); however, if $q \notin \myset{S}$, there is a small probability that the answer is incorrect (a false positive).
This probabilistic response typically yields significant savings in memory -- the total size of a BF can be much smaller than the combined size of all items it encodes. We consider two variants of BFs, described in the following.

\textbf{Non-partitioned BF.} In this case, the BF representation of $\myset{S}$ is a bit vector $\myvector{b} \in \{ 0, 1\}^{L_{np}}$, initialized to $\myvector{b} = (0, 0, ..., 0)$.
The number of bits that are used is $B_{np}=L_{np}$.
Hash functions $h_1, h_2, ..., h_M$, with $h_m: \myset{U} \rightarrow \{1, 2, ..., L_{np}\} \forall m$, map an item to a single bit of $\myvector{b}$. 
To insert a database item $x \in \myset{S}$ into the BF, we hash it $M$ times and the bits $\myvector{b}[h_1(x)]$, $\myvector{b}[h_2(x)]$, $...$, $\myvector{b}[h_M(x)]$ are set to $1$.
This repeats for each database item, so more and more bits are set.
Insertion of additional items is simple, but deletion is not possible.
At query time, the BF responds that $q \in \myset{S}$ if $\myvector{b}[h_1(q)] =$ $\myvector{b}[h_2(q)] =$ $...$ $= \myvector{b}[h_M(q)] = 1$, and $q \notin \myset{S}$ otherwise.

\textbf{Partitioned BF.} In this variant, the bit vector $\myvector{b}$ is partitioned into M equal parts $\myvector{b}_m$, each of length $L_p$.
Each hash function $h_m$ only produces bits in its respective partition $\myvector{b}_m$.
The total number of bits is $B_{p}=L_p\times M$.
If $L_p=\frac{L_{np}}{M}$ (which leads to $B_{p}=B_{np}$), the false positive rate is asymptotically the same for partitioned and non-partitioned BFs.



\begin{figure}[t]
 \centering   
\includegraphics{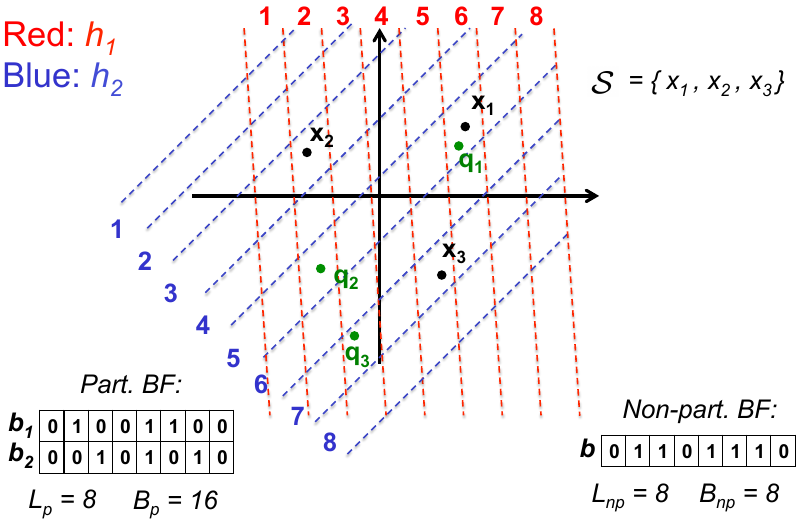}
 \caption{Illustration of a Bloom filter encoding set $\myset{S} = \{ x_1, x_2, x_3 \}$ in 2D. Two hash functions are shown ($M=2$), in red and in blue, with bin numbers marked near the corresponding regions. Partitioned (left) and non-partitioned (right) BFs are presented. Examples of queries are shown in green. Consider that the BF should indicate $q \in \myset{S}$ if the query is close to a database item. Both partitioned and non-partitioned BFs indicate $q_1 \in \myset{S}$ (True Positive) and $q_2 \in \myset{S}$ (False Positive). For $q_3$, the non-partitioned BF indicates $q_3 \in \myset{S}$ (False Positive), and the partitioned BF indicates $q_3 \notin \myset{S}$ (True Negative).}
  \label{fig:BF_illustration}
\end{figure}

\textbf{Distance-sensitive BF.} The BF introduced by \cite{Bloom1970} is designed to decide for the presence of an \textit{exact} match in a database set.
In general retrieval problems, the notion of approximate set membership queries might be more useful.
Such queries are concerned with the question ``is $q$ near an item of $\myset{S}$?''.
For example, if we model a scene as a set and a frame as its item, a query image will unlikely be exactly the same as a frame, and a match may never occur.
We want to find scenes that contain frames which are \textit{similar} to the query image.
Our application is thus more suitable to distance-sensitive Bloom filters (DSBF) \cite{Kirsch2006}, which address this problem, illustrated in \figref{fig:BF_illustration}.
DSBFs are similar to standard BFs, but they are coupled to locality-sensitive hashes (LSH) -- since in this case the hashes must map similar items to the same hash bucket with high probability.

\textbf{BFs and hashing.} As discussed in \cite{Broder2004}, BFs can be seen as a generalization of hashing, supporting more effective trade-offs between number of bits and false positive rate.
In general, BFs obtain lower memory requirements than simple hashing schemes, given a target false positive rate.

\subsection{BF-GD: Using Global Descriptors} \label{subsec:bf_global}
First, we apply the BF framework to our problem in a straightforward way: query images are directly modeled as items, and database scenes as sets of video frames.
For each scene, the constituent frames are hashed into a BF.
A query image can then be matched against the BF of each scene.

To represent query images and video frames, we use global image descriptors -- this method is denoted BF-GD.
We use Fisher vectors (FV) \cite{Jegou2012}, a common technique in multimedia retrieval, whose variants obtain high performance in many datasets \cite{shi2015early,tolias2015image}.
Note that this system discards the ordering of frames: the representation for a given scene is the same regardless of the ordering of its constituent frames.
This is akin to the use of bag-of-words in image retrieval, where the representation is the same regardless of where local features appear in an image.

\subsection{BF-PI: Using Point-Indexed Descriptors} \label{subsec:bf_local}
In this section, we consider a different configuration of the BF framework.
The motivation arises from noticing the two levels of aggregation at play when using BF-GD: local descriptors are first aggregated into FVs per frame, then FVs are aggregated per scene.
It is not clear the impact of these two stages to the discriminativeness of the final scene descriptor.
This leads us to remove one aggregation step, and directly encode Fisher-embedded local features into BFs (they are not aggregated per frame before hashing, as in BF-GD).
Tao \etal \cite{TaoCVPR2014} showed how a FV can be decomposed into the Fisher embedding of each local feature, leading to a point-indexed representation: instead of storing a FV, the database stores an embedded version of each local feature.

The proposed technique is called BF-PI.
Consider a local feature $\myvector{x}$ and a FV with parameters $\{w_k, \mu_k, \sigma_k, k=1\ldots K\}$ denoting the mixture probabilities, mean vectors and diagonal covariance matrices.
As in \cite{TaoCVPR2014}, we employ the point-indexed representation of $\myvector{x}$ using only the Gaussian from the FV which obtains the strongest soft-assignment probability.
The point-indexed representation for $\myvector{x}$ is a triplet:
\begin{align}
\{r; \frac{\gamma_\myvector{x}(r)}{\sqrt{w_r}}; \myvector{d_x} = \sigma_r^{-1}(\myvector{x} - \mu_r)\} \label{eq:pi}
\end{align}
where $r$ is the index of the Gaussian with strongest soft-assignment probability for $\myvector{x}$, $\gamma_\myvector{x}(r)$ is the value of that soft-assignment probability, and $\myvector{d_x}$ is the scaled residual vector between $\myvector{x}$ and the $r$-th Gaussian.
With $\myvector{x}$ represented in this manner, the bucket $h_r(\myvector{d_x})$ in the BF is set to $1$.


\vspace{-10pt}
\subsection{Hash Functions \& Scoring} \label{subsec:hash_fn}
\textbf{LSH families.} 
We consider three LSH families.
The standard metric for comparing FVs is cosine similarity, so a natural choice for this problem is the LSH family for cosine distance \cite{charikar2002similarity}, which uses random hyperplanes -- referred to as LSH-C.
A second family of functions, denoted LSH-S, is a special case of LSH-C, where the components of random hyperplanes are 
either $+1$ or $-1$, picked at random.
This family has been widely used in information retrieval \cite{henzinger2006finding,rajaraman2014}.
We also consider the LSH family for Hamming distance, denoted LSH-B.
This function samples a bit from a binarized signature, and can be generalized to real-valued vectors by using random axis-aligned hyperplanes.
In practice, we want to map each item to $L$ buckets.
To accomplish that, each of the $M$ hash functions we use is composed of $n$ hyperplanes, thus mapping each item to one out of $2^n = L$ buckets.

\textbf{Domain of hash functions.}
The natural choice for the domain of hash functions is the original space where items lie.
We denote hash functions of this type as vector-based hashes (VBH).
For a FV with $K$ Gaussians, and local descriptors having $d$ dimensions, FVs lie in $\mathcal{R}^{K\times d}$.
Thus, in the BF-GD case, $h_{VBH}:\mathcal{R}^{K\times d} \to 2^n$.
Another possibility is to divide FVs into chunks corresponding to their Gaussians, and hash each chunk separately.
We denote hash functions of this type as Gaussian-based hashes (GBH), $h_{GBH}:\mathcal{R}^{d} \to 2^n$.
In the BF-PI version, we are interested in hashing $d$-dimensional point-indexed descriptors into $2^n$ buckets.
Thus, GBH is also applicable to this version.

\textbf{Quantizer-based hashing.}
Recent work shows that quantization outperforms random hashes for approximate nearest neighbor tasks \cite{balu2014beyond,pauleve2010locality}.
With that motivation, we employ $K$-means to construct a vector quantizer (VQ), and use it as a hash function: an item is inserted into the bucket corresponding to the centroid it is closest to.


\textbf{Scoring.}
At query time, the query image is processed in the same way video frames are processed at indexing time.
To score scenes, we explore two techniques.
We restrict the presentation to the case of BF-GD, using a non-partitioned BF (scoring for other configurations is similar).
First, we consider scoring based on the number of hash matches ($S^{\#}$).
Given the query image descriptor $\myvector{q}$ and the $m$-th hash function, the score $S^{\#}_v$ of database scene $v$ is updated as:
\begin{align} 
S_v^{\#} : = S_v^{\#} + \myvector{b}_v[h_m(\myvector{q})]
\end{align} 
In other words, the score of scene $v$ is incremented if its $h_m(\myvector{q})$-th bucket is set.
Another option is to use TF-IDF, as is common in information retrieval:
for the same case as above, the score $S^T_v$ of scene $v$ can be computed as:
\begin{align} 
S_v^{T} : = S_v^{T} + \myvector{b}_v[h_m(\myvector{q})] \cdot \frac{w_{h_m(\myvector{q})}^2}{(\sum_{l} \myvector{b}_v[l] w_{l}^2)^\alpha}
\end{align} 
where $w_l$ corresponds to the IDF weight of bucket $l$ and $(\sum_{l} \myvector{b}_v[l] w_{l}^2)^\alpha$ denotes a normalization factor, where $\alpha$ is empirically chosen ($\alpha=0.5$ corresponds to $L_2$ normalization).

\section{Experiments}  \label{sec:experiments}

\textbf{Datasets.}
Our experiments use 3 datasets (2 of them introduced in this work).
The existing Stanford I2V (SI2V) dataset is the largest for this research problem \cite{Araujo2015}. It contains news videos, and query images are collected from the web.
The new Video Bookmarking (VB) dataset uses the same videos as SI2V, but the queries contain displays with a frame of a video being played.
This models the case where a user wants to retrieve the video being played, e.g., to resume playback in a different device.
The third dataset, also introduced in this work, contains lecture videos (ClassX), with queries being clean images of slides.
In all cases, we extract $1$ frame per second.
Extensive experiments are conducted on reduced dataset versions: SI2V-600k, VB-600k and ClassX-600k (each containing 600k frames and 160 hours of video).
Large-scale experiments use SI2V-4M, VB-4M (4M frames and 1,079 hours of video) and ClassX-1.5M (1.5M frames and 408 hours of video).
More than 200 queries are used per dataset.
To train auxiliary structures (e.g., GMM, PCA), we use independent datasets.
All datasets are presented in greater detail in Appendices A and B.

\textbf{Local and global descriptors.} 
For local descriptor extraction, we detect Hessian-affine keypoints \cite{mikolajczyk2004scale} and describe them using SIFT \cite{Lowe2004}.
Using PCA, the dimensionality of SIFT descriptors is reduced to $d=32$.
For FVs \cite{Jegou2012} and point-indexed FVs \cite{TaoCVPR2014}, we use $K=512$ Gaussians.

\textbf{BF parameters.}
We set $M=K=512$, which is experimentally shown as a good choice.
We vary $n$, the number of bits obtained per hash function.
For a given $n$, an item can be mapped into $2^n$ buckets in the BF.
For TF-IDF scoring, we experiment with $\alpha \in \{0, 0.25, 0.5, 0.75, 1\}$ in the 600k datasets.
We tune $n$ and $\alpha$ in the smaller-scale experiments, and use their optimal values for large-scale experiments.

\textbf{Performance assessment.} 
We follow the evaluation procedure from previous work \cite{Araujo2015ICIP,Araujo2015}, to obtain comparable numbers: results are evaluated using mean Average Precision (mAP).
For configurations which use LSH functions, we run the experiments 10 times, and report the averaged mAP.

\begin{figure}[t]
 \centering   
\includegraphics{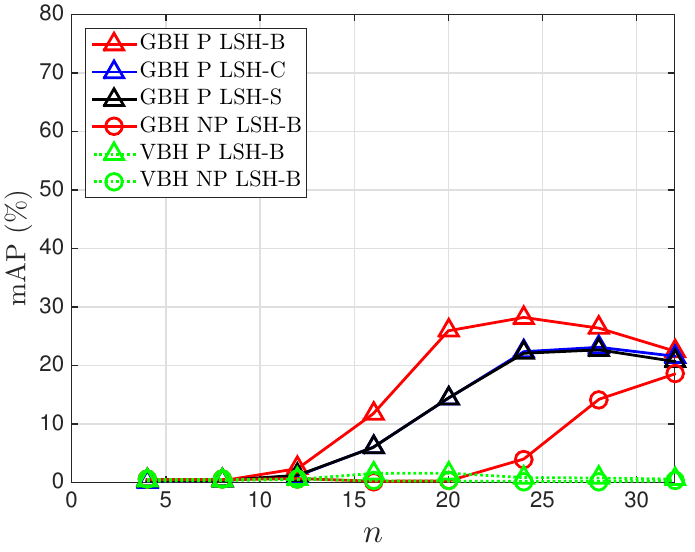}
\label{fig:global_bitsampling_vary_p_np_gbh_vbh}
\vspace{-4mm}
 \caption{BF-GD retrieval results using the SI2V-600k dataset: mAP as a function of $n$. All curves use scoring based on the number of hash matches. Comparison of partitioned (P) versus non-partitioned (NP) BFs; GBH versus VBH hashes; LSH-B versus LSH-C and LSH-S.}
 \label{fig:bf_gd}
\end{figure}

\textbf{Results: BF-GD.}
\figref{fig:bf_gd} presents BF-GD results in the SI2V-600k dataset.
First, note that GBH outperforms VBH significantly: this can be understood since FVs aggregate different types of visual information per Gaussian, and the correlation between different Gaussians might be weak.
\figref{fig:bf_gd} also compares partitioned (P) and non-partitioned (NP) BFs.
For a fair comparison, we should have $B_p = B_{np}$: P-BF using $M = 512 = 2^9$ bit vectors of length $2^n$ should be compared to NP-BF using a bit vector of length $2^{n+9}$.
In this case, P-BF outperforms NP-BF.
Finally, \figref{fig:bf_gd} shows that LSH-B outperforms LSH-C and LSH-S.
This might be surprising, as LSH-B is much simpler than LSH-C and LSH-S.
However, these results agree with previous work \cite{Perronnin2010} which shows that this simple technique outperforms spectral hashing \cite{weiss2008spectral} and LSH-C for image retrieval.
Overall, though, BF-GD obtains limited mAP, showing that a straightforward BF aggregation method may not be the best choice for this problem.

\setlength{\belowcaptionskip}{-5pt} 

\setlength{\belowcaptionskip}{-10pt} 

\begin{table}[!t]
\scriptsize
\centering
\begin{tabular}{l c c c}
  \toprule
        & SI2V-600k & VB-600k & ClassX-600k \\
  \toprule
Scene FV$^\star$ (DoG) \cite{Araujo2015ICIP} & 47.33 & - & - \\
Scene FV$^\star$ & 50.01 & 62.17 & \bf{66.71} \\
BF-GD LSH-B & 35.73 & 38.53 & 36.02 \\
BF-PI LSH-B & 70.46 & 65.76 & 45.05 \\
BF-PI VQ & \bf{73.75} & \bf{67.67} & 63.66 \\
  \bottomrule
\end{tabular}
\vspace{-2mm}
\caption{Summary of retrieval results (mAP in \%) for the 600k datasets. All techniques use Hessian-affine keypoints, except for \cite{Araujo2015ICIP}, which uses difference-of-Gaussian (DoG) keypoints. The BF techniques presented here use GBH hashes, partitioned BFs and TF-IDF.}
  \label{tab:ss_results}
\end{table}

\begin{figure}[t]
 \centering   
\includegraphics{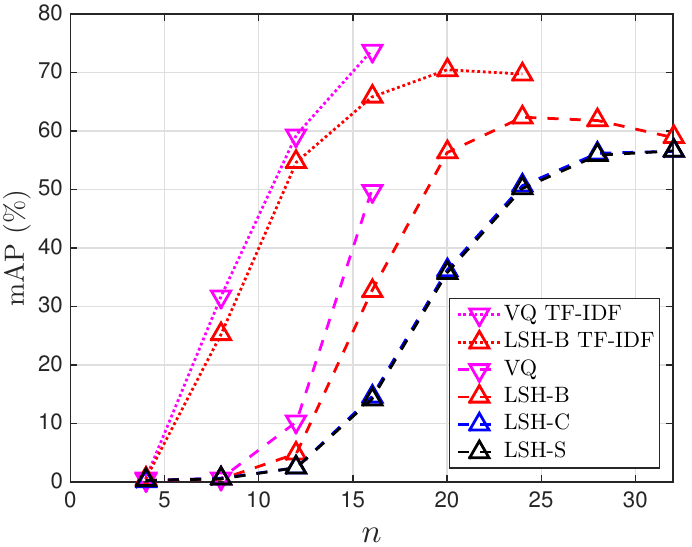}
\vspace{-4mm}
 \caption{BF-PI retrieval results using the SI2V-600k dataset: mAP as a function of $n$. All curves use partitioned BFs and GBH. Comparison of VQ versus LSH hashes, and different scoring techniques.}
  \label{fig:bf_pi}
\end{figure}

\textbf{Results: BF-PI.}
\figref{fig:bf_pi} compares the different hashing and scoring techniques, when using BF-PI.
BF-PI provides a substantial improvement in mAP, compared to BF-GD, of more than $30\%$.
This demonstrates the benefit of removing the aggregation per frame before hashing.
This figure shows that LSH-B outperforms LSH-C and LSH-S also when BF-PI is used.
\figref{fig:bf_pi} further introduces results using the TF-IDF scoring method and VQ hashes.
In this case, we use $n\leq 16$ to limit memory and computational complexity.
VQ-based hashing improves retrieval performance compared to the scheme that makes use of LSH-B.
BF-PI using $n=16$, coupled with VQ-based hashing and TF-IDF scoring, outperforms all other BF configurations we experimented with.

\textbf{Comparison to state-of-the-art.}
\tabref{tab:ss_results} presents summarized results for experiments on the 600k datasets.
We compare the proposed methods against the state-of-the-art in scene-based signatures: Scene FV$^\star$ \cite{Araujo2015ICIP} (binarized FVs with $2048$ Gaussians), which used difference-of-Gaussian keypoints, on the SI2V dataset (we use their public code to reproduce these results).
We also implement Scene FV$^\star$ using Hessian-affine keypoints (which boosts performance of \cite{Araujo2015ICIP}).
The proposed BF-PI scheme, coupled with VQ hashing, outperforms other approaches significantly for SI2V-600k, by $26\%$ mAP compared to \cite{Araujo2015ICIP}.
In the VB-600k dataset, it also outperforms other approaches, but with a smaller margin: $5.50\%$ better than Scene FV$^\star$.
In the ClassX-600k dataset, BF-PI VQ is slightly worse than Scene FV$^\star$, by $3.05\%$.

\setlength{\belowcaptionskip}{-5pt} 

\setlength{\belowcaptionskip}{-10pt} 

\begin{table}[t]
\scriptsize
\centering
\begin{tabular}{l c c c}
  \toprule
        & mAP (\%) & Latency (secs) & Memory (GB) \\
  \toprule
  \multicolumn{2}{l}{\bf{SI2V-4M dataset}} & & \\
Frame FV$^\star$ & 72.44 & 0.4118 & 20.59\\
Scene FV$^\star$ \cite{Araujo2015ICIP} & 49.71 & 0.1643 & \bf{3.01} \\
BF-PI VQ (ours) & \bf{74.08} & \bf{0.0431} & 10.76 \\
    \midrule
  \multicolumn{2}{l}{\bf{VB-4M dataset}} & & \\
Frame FV$^\star$ & 75.97 & 0.4423 & 20.59\\
Scene FV$^\star$ \cite{Araujo2015ICIP} & 67.37 & 0.2106 & \bf{3.01} \\
BF-PI VQ (ours) & \bf{76.25} & \bf{0.1101} & 10.76 \\
    \midrule
  \multicolumn{2}{l}{\bf{ClassX-1.5M dataset}} & & \\
Frame FV$^\star$ & 64.21 & 0.1984 & 7.67\\
Scene FV$^\star$ \cite{Araujo2015ICIP} & 64.47 & 0.0365 & \bf{0.42} \\
BF-PI VQ (ours) & \bf{67.60} & \bf{0.0357} & 1.20 \\
  \bottomrule
\end{tabular}
\vspace{-2mm}
\caption{Summarized results for large-scale experiments, comparing the proposed BF-PI technique against the previous state-of-the-art \cite{Araujo2015ICIP} and against a reference frame-based technique. All methods use inverted index structures and Hessian-affine keypoints. Retrieval latency results are per query, using one core on an Intel Xeon 2.4GHz.}
  \label{tab:ls_results}
\end{table}

\textbf{Large-scale experiments} 
compare our best method to the state-of-the-art \cite{Araujo2015ICIP} (using Hessian-affine keypoints, since it improves retrieval performance), and to a technique which indexes every frame in the database using binarized FVs with 512 Gaussians, denoted Frame FV$^\star$.
Results are presented in \tabref{tab:ls_results}, including latency and memory figures.
Contrary to previous work \cite{Araujo2015ICIP}, we use inverted index retrieval structures for all techniques, to obtain practical large-scale systems.
For Scene FV$^\star$ and Frame FV$^\star$, we use the Multi-Block Indexing Table (MBIT) \cite{duan2016overview,wang2014component} method, which was introduced to provide speedup over simple linear search for binarized FVs (we used MBIT's source code, which is available online).
For BF-PI, we can represent it in an inverted index format, which is straightforward.
For BF-PI and Scene FV$^\star$, we re-rank the top scene results using shot-based FV$^\star$s, as in \cite{Araujo2015ICIP}.
It can be seen that BF-PI is much faster (up to $4\times$) and much more accurate (up to $24\%$) than Scene FV$^\star$, in all datasets.
BF-PI also outperforms Frame FV$^\star$: up to 10$\times$ faster retrieval with slightly improved mAP.
This demonstrates the improved scalability of our technique, compared to the common approach of frame-based indexing, which has been widely used for query-by-image video retrieval.

\section{Conclusion} \label{sec:conclusion}

In this work, we explore aggregation of visual information from scenes into Bloom filters, enabling efficient and effective video retrieval using image queries.
First, we show that a straightforward application of Bloom filters to our problem, using global image descriptors, obtains limited retrieval accuracy.
Our best-performing scheme adapts the Bloom filter framework: the key is to hash discriminative local descriptors into scene-based signatures.
The techniques are evaluated by considering different hash functions and score computation methods.
Large-scale experiments show that our system achieves high retrieval accuracy and reduced query latency in three datasets, outperforming the previous state-of-the-art.

\begin{appendices}
\section*{Appendix A: Datasets for Query-by-Image \\ Video Retrieval} \label{sec:modeling}

\begin{figure*}[!t]
\centering
\includegraphics{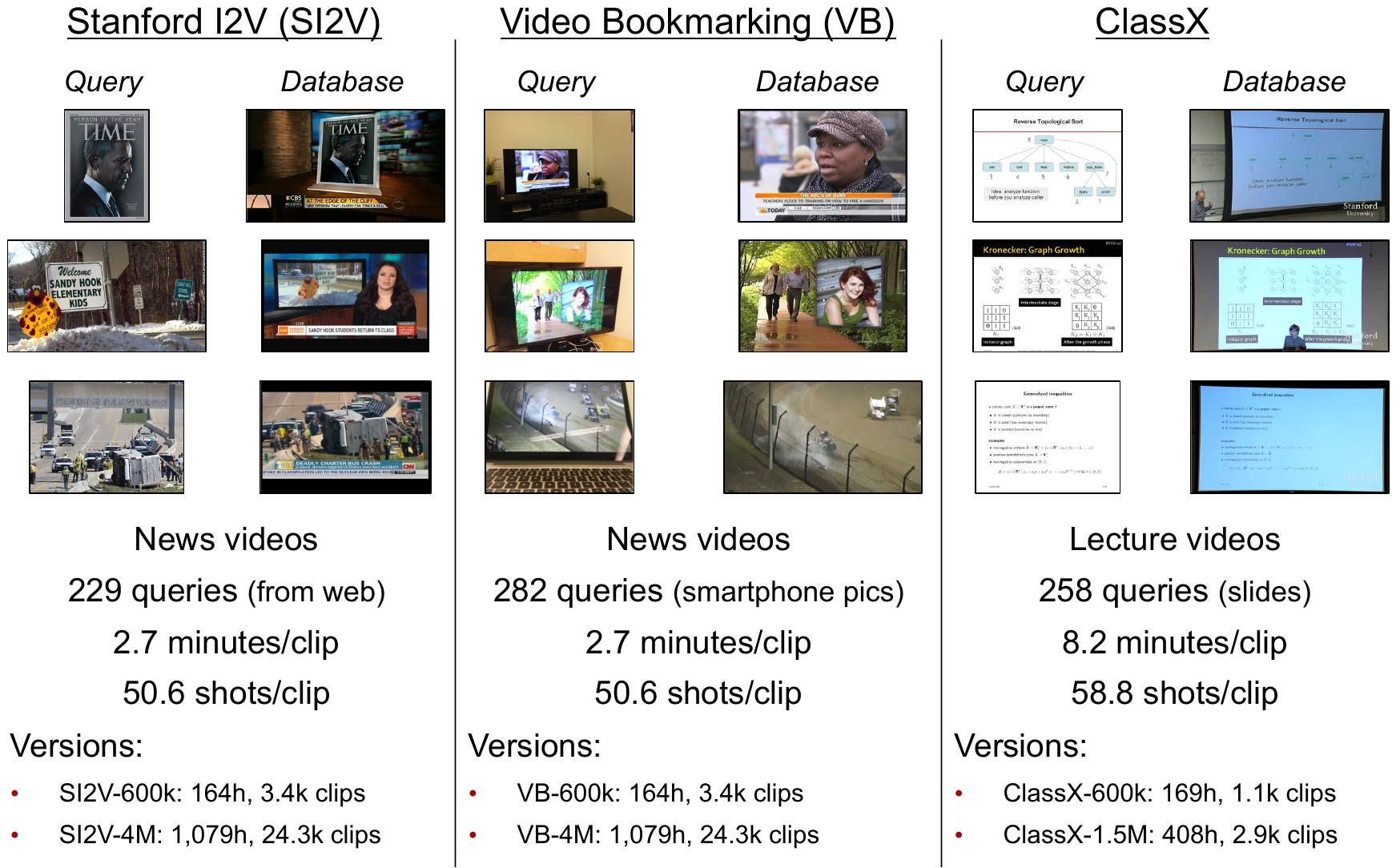}
\caption{Illustration of the query-by-image video retrieval datasets used in this work. Left: Stanford I2V (SI2V) dataset. Center: Video Bookmarking (VB) dataset. Right: ClassX dataset.}
\label{fig:i2v_datasets}
\end{figure*}

This section presents the large-scale query-by-image video retrieval datasets used in this paper.
They are illustrated in \figref{fig:i2v_datasets}.
Two new datasets are introduced: Video Bookmarking and ClassX.
They are released to the community, to encourage comparative experiments.

\subsubsection*{Stanford I2V (SI2V)} \label{sec:stanfordi2v}

This dataset was introduced in \cite{Araujo2015}, and is publicly available since 2015, at \url{https://purl.stanford.edu/zx935qw7203}.
This is the largest available dataset for query-by-image video retrieval experiments.
The video clips correspond to scenes: in this case, they are news stories from 39 recurring newscasts, and the query set is composed of 229 images collected from the web, with annotated ground-truth.
For example, a video clip could start with an anchor in a studio, then it would transition to a reporter in the field, then original footage of an event would be presented, then the story would be wrapped up by the anchor in the studio.
The clips are on average 2.7 minutes long and contain on average 50.6 different shots.
In this work, we experiment with two versions of this dataset.
In both versions, the full query set is used.
The dataset versions are: (i) SI2V-600k: a version of the dataset containing 3,401 clips, with 164 hours of video and 589,965 frames extracted at 1 frame per second (fps); (ii) SI2V-4M: a large-scale version of the dataset containing 24,317 clips, with 1,079 hours of video and 3,972,602 frames extracted at 1fps.

\subsubsection*{Video Bookmarking (VB)} \label{sec:vbookmarking}

This new dataset makes use of the same database videos as the SI2V dataset, but it introduces a different query set.
It is available at \url{https://purl.stanford.edu/zx935qw7203}.
The application modeled by this dataset corresponds to the case where a user is interested in retrieving the video being played, e.g., to resume playback in a different device.
The queries contain displays (from TVs or computers) with a frame of a video being played.
Smartphone and tablet cameras are used to generate these queries, and the number of queries in this dataset is 282.
The VB-600k and VB-4M versions of this dataset are based on the same database videos as the SI2V dataset.
In both versions, the full query set is used.

\subsubsection*{ClassX} \label{sec:classx}

This new dataset contains lecture video segments, which have been collected from 21 engineering-related university courses, in 2013 and 2014.
It is available at \url{http://purl.stanford.edu/sf888mq5505}.
These lecture segments have been collected as part of the ClassX system \cite{ClassX} at Stanford University.
Each database video clip corresponds to a lecture segment that introduces a single topic (i.e., each clip corresponds to a scene for this type of content).
For example, a video clip would present the notion of locality-sensitive hashing, in the context of a lecture that presents different techniques to mine large datasets of documents.
The clips are on average 8.2 minutes long and contain on average 58.8 different shots.
The queries are 258 clean images of slides which are shown during a particular lecture.
In this work, we experiment with two versions of this dataset.
In both versions, the full query set is used.
The dataset versions are: (i) ClassX-600k: a version of the dataset containing 1,166 clips, with 169 hours of video and 607,747 frames extracted at 1fps; (ii) ClassX-1.5M: a large-scale version of the dataset containing 2,981 clips, with 408 hours of video and 1,478,978 frames extracted at 1fps.

\subsubsection*{Discussion}
These three datasets present very different characteristics.
SI2V and VB present much larger variety of content than ClassX -- the average number of shots per clip is similar, but the average clip duration is much longer for the ClassX dataset.
Regarding queries, in the ClassX dataset the matching object covers the entire query image, while this is not usually the case for the SI2V and VB datasets.
In SI2V, the database ground-truth matching information usually occupies only a fraction of the total frame, while in the other two datasets usually it covers the entire frame.

\section*{Appendix B: Training Datasets} \label{sec:modeling}

This section presents the independent datasets which were used for training auxiliary retrieval structures in this paper.
Two new datasets are introduced: ClassX-Training and Slideshare-1M.
They are released to the community, to encourage comparative experiments.

\subsubsection*{Flickr60k} \label{subsec:flickr60k}

This standard dataset (introduced by J\'{e}gou \etal \cite{Jegou2008}) contains 67,714 images, provided together with 140 million SIFT \cite{Lowe2004} descriptors, extracted from keypoints detected using the Hessian-Affine keypoint detector \cite{mikolajczyk2004scale}.
These data were used to train Gaussian mixture models and principal component analysis for the SI2V and VB datasets.

\subsubsection*{ClassX-Training} \label{subsec:classx_training}

This new dataset has been used to train Gaussian mixture models and principal component analysis used for retrieval with the ClassX dataset.
The data is composed of slides for 12 engineering-related courses, different from the ones used in the ClassX dataset.
A total of $8,619$ slides compose this training dataset.
It is available at \url{http://purl.stanford.edu/sf888mq5505}.

\subsubsection*{MIRFLICKR-1M} \label{subsec:flickr1m}

This dataset contains 1 million images downloaded from Flickr under the creative commons license \cite{huiskes08}.
It has been used to train quantizers for the Bloom filter technique, when performing retrieval on the SI2V and VB datasets.

\subsubsection*{Slideshare-1M} \label{subsec:slideshare}

This new dataset, which contains approximately 1 million slides, was used for training quantizers for the Bloom filter technique, when performing retrieval on the ClassX dataset.
977,605 slides (from 31,923 slide decks) were collected from Slideshare using the available API, with tags related to engineering and science, as the ClassX dataset (which is the target dataset in this case) contains slides mostly related to these topics.
It is available at \url{http://purl.stanford.edu/mv327tb8364}.

\end{appendices}

%
\bibliographystyle{abbrv}
{\small
\bibliography{literature/mm16_refs}}  

\begin{thebibliography}{10}

\bibitem{Araujo2015ICIP}
A.~Araujo, J.~Chaves, R.~Angst, and B.~Girod.
\newblock {Temporal Aggregation for Large-Scale Query-by-Image Video
  Retrieval}.
\newblock In {\em Proc. ICIP}, 2015.

\bibitem{Araujo2015}
A.~Araujo, J.~Chaves, D.~Chen, R.~Angst, and B.~Girod.
\newblock {Stanford I2V: A News Video Dataset for Query-by-Image Experiments}.
\newblock In {\em Proc. ACM MMSys}, 2015.

\bibitem{ballan2015data}
L.~Ballan, M.~Bertini, G.~Serra, and A.~Del~Bimbo.
\newblock {A Data-Driven Approach for Tag Refinement and Localization in Web
  Videos}.
\newblock {\em Computer Vision and Image Understanding}, 140:58--67, 2015.

\bibitem{Ballas2013}
N.~Ballas, M.~Redi, A.~Hamadi, B.~Gao, B.~Labbe, B.~Merialdo, C.~Zhu, L.~Chen,
  B.~Safadi, N.~Derbas, Y.~Tang, A.~Benoit, H.~L. Borgne, B.~Mansencal,
  E.~Dellandrea, J.~Benois-Pineau, P.~Lambert, P.~Gosselin, M.~Budnik,
  T.~Strat, D.~Picard, S.~Ayache, G.~Quenot, and C.~Bichot.
\newblock {IRIM at TRECVID 2014: Semantic Indexing and Instance Search}.
\newblock In {\em Proc. TRECVID}, 2014.

\bibitem{balu2014beyond}
R.~Balu, T.~Furon, and H.~J{\'e}gou.
\newblock {Beyond Project and Sign for Cosine Estimation with Binary Codes}.
\newblock In {\em Proc. ICASSP}, 2014.

\bibitem{Bloom1970}
B.~Bloom.
\newblock {Space/Time Trade-offs in Hash Coding with Allowable Errors}.
\newblock {\em Communications of the ACM}, 13, 1970.

\bibitem{Broder2004}
A.~Broder and M.~Mitzenmacher.
\newblock {Network Applications of Bloom Filters: A Survey}.
\newblock {\em Internet Mathematics}, 2004.

\bibitem{charikar2002similarity}
M.~S. Charikar.
\newblock {Similarity Estimation Techniques from Rounding Algorithms}.
\newblock In {\em Proc. ACM Symposium on Theory of Computing}, 2002.

\bibitem{chen2014event}
J.~Chen, Y.~Cui, G.~Ye, D.~Liu, and S.-F. Chang.
\newblock {Event-Driven Semantic Concept Discovery by Exploiting Weakly Tagged
  Internet Images}.
\newblock In {\em Proc. ICMR}, 2014.

\bibitem{ClassX}
{ClassX}.
\newblock \url{http://classx.stanford.edu}. Accessed Apr. 2016.

\bibitem{duan2012exploiting}
L.~Duan, D.~Xu, and S.-F. Chang.
\newblock {Exploiting Web Images for Event Recognition in Consumer Videos: A
  Multiple Source Domain Adaptation Approach}.
\newblock In {\em Proc. CVPR}, 2012.

\bibitem{duan2016overview}
L.-Y. Duan, V.~Chandrasekhar, J.~Chen, J.~Lin, Z.~Wang, T.~Huang, B.~Girod, and
  W.~Gao.
\newblock {Overview of the MPEG-CDVS Standard}.
\newblock {\em IEEE Transactions on Image Processing}, 25(1), 2016.

\bibitem{henzinger2006finding}
M.~Henzinger.
\newblock {Finding Near-Duplicate Web Pages: A Large-Scale Evaluation of
  Algorithms}.
\newblock In {\em Proc. ACM SIGIR}. ACM, 2006.

\bibitem{jain2015objects2action}
M.~Jain, J.~C. van Gemert, T.~Mensink, and C.~G. Snoek.
\newblock {Objects2action: Classifying and Localizing Actions Without any Video
  Example}.
\newblock In {\em Proc. ICCV}, 2015.

\bibitem{Jegou2008}
H.~J\'{e}gou, M.~Douze, and C.~Schmid.
\newblock {Hamming Embedding and Weak Geometric Consistency for Large Scale
  Image Search}.
\newblock In {\em Proc. ECCV}. Springer, 2008.

\bibitem{Jegou2012}
H.~J\'{e}gou, F.~Perronnin, M.~Douze, J.~Sanchez, P.~P{\'{e}}rez, and
  C.~Schmid.
\newblock {Aggregating Local Image Descriptors into Compact Codes}.
\newblock {\em IEEE Transactions on Pattern Analysis and Machine Intelligence},
  34(9), 2012.

\bibitem{Kirsch2006}
A.~Kirsch and M.~Mitzenmacher.
\newblock {Distance-Sensitive Bloom Filters}.
\newblock In {\em Proc. Workshop on Algorithm Engineering and Experiments
  (ALENEX)}, 2006.

\bibitem{le2011trecvid}
D.-D. Le, C.-Z. Zhu, S.~Poullot, V.~Q. Lam, D.~A. Duong, and S.~Satoh.
\newblock {National Institute of Informatics, Japan at TRECVID 2011}.
\newblock In {\em Proc. TRECVID}, 2011.

\bibitem{rajaraman2014}
J.~Leskovec, A.~Rajaraman, and J.~Ullman.
\newblock {\em {Mining of Massive Datasets}}.
\newblock Cambridge University Press, 2014.

\bibitem{Lowe2004}
D.~Lowe.
\newblock {Distinctive Image Features from Scale-Invariant Keypoints}.
\newblock {\em International Journal of Computer Vision}, 60(2), 2004.

\bibitem{huiskes08}
B.~T. M.~Huiskes and M.~Lew.
\newblock {New Trends and Ideas in Visual Concept Detection: The MIR Flickr
  Retrieval Evaluation Initiative}.
\newblock In {\em Proc. ICMR}, 2010.

\bibitem{meng2016object}
J.~Meng, J.~Yuan, J.~Yang, G.~Wang, and Y.-P. Tan.
\newblock {Object Instance Search in Videos via Spatio-Temporal Trajectory
  Discovery}.
\newblock {\em IEEE Transactions on Multimedia}, 18(1), 2016.

\bibitem{merler2012semantic}
M.~Merler, B.~Huang, L.~Xie, G.~Hua, and A.~Natsev.
\newblock {Semantic Model Vectors for Complex Video Event Recognition}.
\newblock {\em IEEE Transactions on Multimedia}, 14(1), 2012.

\bibitem{mikolajczyk2004scale}
K.~Mikolajczyk and C.~Schmid.
\newblock {Scale \& Affine Invariant Interest Point Detectors}.
\newblock {\em {International Journal of Computer Vision}}, 60(1), 2004.

\bibitem{2015trecvidover}
P.~Over, G.~Awad, J.~Fiscus, M.~Michel, D.~Joy, A.~Smeaton, W.~Kraaij,
  G.~Quenot, and R.~Ordelman.
\newblock {TRECVID 2015 -- An Overview of the Goals, Tasks, Data, Evaluation
  Mechanisms and Metrics}.
\newblock In {\em Proc. TRECVID}, 2015.

\bibitem{pauleve2010locality}
L.~Paulev{\'e}, H.~J{\'e}gou, and L.~Amsaleg.
\newblock {Locality Sensitive Hashing: A Comparison of Hash Function Types and
  Querying Mechanisms}.
\newblock {\em {Pattern Recognition Letters}}, 31(11), 2010.

\bibitem{Peng2015}
Y.~Peng, J.~Zhang, X.~Huang, M.~Sun, X.~He, P.~Tang, Y.~Zhao, J.~Zhao, J.~Qi,
  and J.~Zhang.
\newblock {PKU-ICST at TRECVID 2015: Instance Search Task}.
\newblock In {\em Proc. TRECVID}, 2015.

\bibitem{Perronnin2010}
F.~Perronnin, Y.~Liu, J.~Sanchez, and H.~Poirier.
\newblock {Large-Scale Image Retrieval with Compressed Fisher vectors}.
\newblock In {\em Proc. CVPR}, 2010.

\bibitem{Rasheed2003}
Z.~Rasheed and M.~Shah.
\newblock {Scene Detection In Hollywood Movies and TV Shows}.
\newblock In {\em Proc. CVPR}, 2003.

\bibitem{shi2015early}
M.~Shi, Y.~Avrithis, and H.~J{\'e}gou.
\newblock {Early Burst Detection for Memory-Efficient Image Retrieval}.
\newblock In {\em Proc. CVPR}, 2015.

\bibitem{Sivic2003}
J.~Sivic and A.~Zisserman.
\newblock {Video Google: A Text Retrieval Approach to Object Matching in
  Videos}.
\newblock In {\em Proc. ICCV}, 2003.

\bibitem{sun2015temporal}
C.~Sun, S.~Shetty, R.~Sukthankar, and R.~Nevatia.
\newblock {Temporal Localization of Fine-Grained Actions in Videos by Domain
  Transfer from Web Images}.
\newblock In {\em Proc. ACM MM}, 2015.

\bibitem{TaoCVPR2014}
R.~Tao, E.~Gavves, C.~G.~M. Snoek, and A.~W.~M. Smeulders.
\newblock {Locality in Generic Instance Search from One Example}.
\newblock In {\em Proc. CVPR}, 2014.

\bibitem{tolias2015image}
G.~Tolias, Y.~Avrithis, and H.~J{\'e}gou.
\newblock {Image Search with Selective Match Kernels: Aggregation Across Single
  and Multiple Images}.
\newblock {\em International Journal of Computer Vision}, 2015.

\bibitem{Truong2003}
B.~Truong, S.~Venkatesh, and C.~Dorai.
\newblock {Scene Extraction in Motion Pictures}.
\newblock {\em IEEE Transactions on Circuits and Systems for Video Technology},
  13(1), 2003.

\bibitem{wang2016learning}
J.~Wang, W.~Liu, S.~Kumar, and S.-F. Chang.
\newblock {Learning to Hash for Indexing Big Data -- A Survey}.
\newblock {\em Proceedings of the IEEE}, 104(1), 2016.

\bibitem{wang2014component}
Z.~Wang, L.-Y. Duan, J.~Lin, T.~Huang, W.~Gao, and M.~Bober.
\newblock {Component Hashing of Variable-Length Binary Aggregated Descriptors
  for Fast Image Search}.
\newblock In {\em Proc. ICIP}, 2014.

\bibitem{weiss2008spectral}
Y.~Weiss, A.~Torralba, and R.~Fergus.
\newblock {Spectral Hashing}.
\newblock In {\em Proc. NIPS}, 2008.

\bibitem{Yeung1998}
M.~Yeung, B.-L. Yeo, and B.~Liu.
\newblock {Segmentation of Video by Clustering and Graph Analysis}.
\newblock {\em Computer Vision and Image Understanding}, 71(1), 1998.

\bibitem{Zhu2014}
C.-Z. Zhu, Y.-H. Huang, and S.~Satoh.
\newblock {Multi-Image Aggregation for Better Visual Object Retrieval}.
\newblock In {\em Proc. ICASSP}, 2014.

\bibitem{Zhu2013}
C.-Z. Zhu, H.~Jegou, and S.~Satoh.
\newblock {Query-Adaptive Asymmetrical Dissimilarities for Visual Object
  Retrieval}.
\newblock In {\em Proc. ICCV}, 2013.

\bibitem{Zhu2012}
C.-Z. Zhu and S.~Satoh.
\newblock {Large Vocabulary Quantization for Searching Instances from Videos}.
\newblock In {\em Proc. ICMR}, 2012.

\end{thebibliography}
%
%
\end{document}